\documentclass[11pt,a4paper]{article}

\usepackage{amsfonts,amscd,amssymb,amsmath,amsthm,latexsym}

\title{Some Cosmological  Solutions of a New Nonlocal Gravity Model}

\small{\author{I. Dimitrijevic$^1$, B. Dragovich$^{2,3}$, A. S. Koshelev$^4$,  Z. Rakic$^1$ and J. Stankovic$^5$ \\
$^1$Faculty of Mathematics, University of Belgrade,  Belgrade, Serbia   \\
$^2$Institute of Physics, University of Belgrade,   Belgrade, Serbia \\
$^3$Mathematical Institute of SASA, Belgrade, Serbia; dragovich@ipb.ac.rs \\
$^4$Departamento  de F\'isica and Centro  de  Matem\'atica  e
Aplica\c c\~oes, \\  Universidade  da  Beira  Interior,   6200  Covilh\~a,
Portugal   \\
$^5$Teacher Education Faculty, University of Belgrade,   Belgrade, Serbia}}

\date{}

\begin{document}

\maketitle


\begin{abstract}
In this paper, we investigate a nonlocal modification of general relativity (GR) with action  $S =  \frac{1}{16\pi G} \int [ R- 2\Lambda + (R-4\Lambda) \, \mathcal{F}(\Box) \, (R-4\Lambda) ] \, \sqrt{-g}\; d^4x ,$ where $\mathcal{F} (\Box) = \sum_{n=1}^{+\infty} f_n \Box^n$  is an analytic function of the d'Alembertian $\Box$. We found a few exact cosmological solutions of the corresponding equations of motion. There are two solutions which are valid only if  $\Lambda \neq 0, \,  k = 0,$ and they have not analogs in Einsten's gravity with cosmological  constant  $\Lambda$. One of these two solutions is $ a (t) = A \, \sqrt{t} \, e^{\frac{\Lambda}{4} t^2} ,$ that mimics  properties similar to an interference between the radiation and the dark energy.  Another solution is a nonsingular bounce one  -- $ a (t) = A  \, e^{\Lambda t^2}$. For these two solutions, some cosmological aspects are discussed.
We also found  explicit form of the nonlocal operator $\mathcal{F}(\Box)$, which satisfies obtained necessary conditions.
\end{abstract}


\section{Introduction}

 General Relativity \cite{wald}, or in other words Einstein theory of gravity,  is recognized as one of the best physical theories -- with beautiful theoretical properties and significant phenomenological achievements. GR very well describes dynamics of the Solar System. It predicted several important phenomena that were confirmed: deflection of light by the Sun,
gravitational light redshift, gravitational waves, gravitational lensing and black holes.

Despite its extraordinary  success, GR should not be viewed as a final theory of gravity. For example, from the standard cosmological model, which assumes applicability  of GR to the universe as a whole,  follows that the universe is approximately made of 68\% of dark energy (DE),  27\% of dark matter (DM) and only  5\% of visible (standard) matter.  However, DE and DM are not yet experimentally detected, and validity of GR at very large cosmic scales is not confirmed.  Even in the case of discovering DM and DE, there is still a sense to look for modifications of GR  that may mimic the same or similar effects as those of DE and DM. Also, cosmological solutions of GR, under rather general properties of the matter, contain singularity at the cosmic time $t = 0.$ In addition to these  mentioned  astrophysical and cosmological problems, there are also some problems that are pure theoretical and come from quantum gravity and string theory. Note also that there is no known  reliable theoretical principle that might show the right direction for valuable extension of GR. Because of all these shortages, there  are many approaches towards  possible  generalization of Einstein theory of gravity (for a review, see  \cite{faraoni,clifton,nojiri,novello,nojiri1}).

 One of the current ways towards modification of GR  is nonlocal modified gravity, see e.g. \cite{deser,woodard,maggiore,biswas1,biswas2,biswas3,biswas5,biswas4,biswas6,dragovich0,koshelev1a,koshelev1b,koshelev1,koshelev2,koshelev3,koshelev4,eliz,koivisto}.
All nonlocal gravity models contain the d'Alembert-Beltrami operator $\Box$, that is  involved mainly in two ways: 1) in the form $\Box^{-n}$ and 2)  as an analytic function
$F (\Box) = \sum_{n=0}^{+\infty} f_n \Box^n$. Models with $\Box^{-n}$ operator are introduced to investigate  the late cosmic time acceleration without its matter origin.  Some  of such models are given by the action
\begin{equation} \label{1.1}
S = \frac{1}{16\pi G} \int \sqrt{-g} \left(R + \mathcal{L}  \right) \, d^4x ,
\end{equation}
where $\mathcal{L} = R\, f(\Box^{-1}R)$ (see, e.g.  \cite{nojiri,woodard,koivisto}), and  $ \mathcal{L} = -\frac{1}{6} m^2 R \Box^{-2} R $ (see \cite{maggiore} and references therein).

 An interesting and promising class of nonlocal gravity models, that have been recently considered,  is given by
\begin{equation} \label{1.2}
S = \frac{1}{16 \pi G}\int_{\mathcal{M}} \sqrt{-g}\, [R - 2\Lambda + P(R)\, \mathcal{F}(\Box)\, Q(R)]\, d^4x ,
\end{equation}
where $\mathcal{M}$ is a pseudo-Riemannian manifold of signature $(1,3)$ with metric $(g_{\mu\nu})$, $\Lambda$ is the cosmological constant, $P(R)$ and $Q(R)$ are some differentiable functions of the Ricci scalar $R$, and $\mathcal{F} (\Box) = \sum_{n=1}^{+\infty} f_n \Box^n$.
Motivation to use this analytic nonlocal operator  comes from  ordinary and $p$-adic string theory (see \cite{dragovich1} and references therein) and observation that some  analytic nonlocal operators
may improve renormalizability in some quantum gravity models, see \cite{stelle,modesto1,modesto2}.
To have better insight into  effects, preliminary investigation of these models is usually without matter.

 Note again that action \eqref{1.2} contains a class  of simple nonlocal extensions of GR, but still in rather general form. Usually researchers start by a particular expression for $P(R)$ and $Q(R)$ as differentiable functions of $R$, while $\mathcal{F}(\Box)$ is treated as an analytic function of operator $\Box$, whose concrete form is not given at the beginning. For given $P(R)$ and $Q(R)$, the next step is derivation of equations of motion for metric tensor $g_{\mu\nu}$. To consider \eqref{1.2} as nonlocal gravity model of interest for cosmology, equations of motion should have  some useful cosmological solutions. Existence of such (usually exact) cosmological solutions requires some restrictions on the function $\mathcal{F} (\Box) = \sum_{n=1}^{+\infty} f_n \Box^n$, i.e. on its coefficients $f_n$, e.g. see \cite{biswas1,biswas2,biswas3,biswas5,biswas4}. Then with these, and perhaps some additional, restrictions there is a possibility to construct the corresponding concrete function
$\mathcal {F} (\Box)$. Since we do not know {\it a priori} function $\mathcal {F} (\Box)$,  this approach is a reasonable
way to get it.

 Concerning action \eqref{1.2}, the most attention has been paid to the simple case when $P(R) = Q(R) = R$, e.g. see \cite{biswas1,biswas2,koshelev1a,koshelev1b,koshelev2,dimitrijevic1,dimitrijevic2,dimitrijevic3,dimitrijevic4,dimitrijevic5,dimitrijevic6,
dimitrijevic7,dimitrijevic8}.  This investigation started in \cite{biswas1,biswas2} by successful attempt to find nonsingular bouncing
solution of the Big Bang singularity problem in standard cosmology. To find appropriate solution of equations of motion, the ansatz  $\Box R = r R + s$ was used, where $r$ and $s$ are parameters that connect the solution and function $\mathcal {F} (\Box)$.   If also in this case cosmological constant $\Lambda = 0$ then it is some kind of nonlocal generalization of the Starobinsky $R^2$ inflation model, whose various properties are studied in \cite{koshelev1,koshelev2}.

Another very intriguing example of the nonlocal gravity  \eqref{1.2} has $P(R) = Q(R) = \sqrt{R - 2\Lambda}$ \cite{PLB}. One of its exact cosmological solutions is $a(t) = A t^{\frac{2}{3}} e^{\frac{\Lambda}{14}t^2} , \, \Lambda \neq 0, \, k=0.$  This solution mimics properties similar to an interplay of the dark matter and the dark energy. Moreover, computed cosmological parameters are in a good agreement with astronomical observations.  It is worth noting that at the first glance appearance of $\sqrt{R - 2 \Lambda}$ in this model may look strange. However, it can be regarded as a natural nonlocal generalization of the standard local Lagrangian $R - 2 \Lambda.$ Namely, one can introduce nonlocality as follows: $R - 2 \Lambda = \sqrt{R - 2 \Lambda} \, \sqrt{R - 2 \Lambda}\ \to \ \sqrt{R - 2 \Lambda}\, [1 + \mathcal {F} (\Box)] \, \sqrt{R - 2 \Lambda}$.

Nonlocal gravity model which we investigate in this paper has $P(R) = Q(R) = R - 4 \Lambda $,  and the action is given explicitly below in \eqref{2.1}. As we will see, one of the exact cosmological solutions is $a(t) = A \sqrt{t} e^{\frac{\Lambda}{4}t^2} , \, \Lambda \neq 0, \, k=0 ,$ which mimics an interplay between radiation and the dark energy.
 Nonlocal term $(R - 4\Lambda)\ \mathcal {F} (\Box) \ (R - 4\Lambda)$ in this model arose in the process of generalization of the above mentioned model with nonlocality $R\ \mathcal {F} (\Box) \ R $. The starting expression was $(R - R_0)\ \mathcal {F} (\Box) \ (R - R_0)$, where $R_0$ is a constant that may lead to some interesting  background solutions.

Section 2 contains derivation of the equations of motion. Section 3 is devoted to the exact cosmological solutions. Some concluding remarks are in section 4.

\section{New Nonlocal Gravity Model}

The action of our nonlocal gravity model is
\begin{equation} \label{2.1}
S =  \frac{1}{16\pi G} \int [ R- 2\Lambda + (R-4\Lambda) \, \mathcal{F}(\Box) \, (R-4\Lambda) ] \, \sqrt{-g}\; d^4x ,
\end{equation}
where  $\mathcal{F}(\Box)= \displaystyle \sum_{n =1}^{\infty} f_{n}\, \Box^{n}$
and $\Box = \nabla_{\mu}\nabla^{\mu}= \frac{1}{\sqrt{-g}}\, \partial_\mu \, (\sqrt{-g}\, g^{\mu\nu}\, \partial_\nu )$ is the corresponding  d'Alembert-Beltrami operator.  In construction of \eqref{2.1} we started from  action
\begin{equation} \label{2.1a}
S =  \frac{1}{16\pi G} \int [ R- 2\Lambda + (R-R_0) \, \mathcal{F}(\Box) \, (R-R_0) ] \, \sqrt{-g}\; d^4x
\end{equation}
 and found that for $R_0 = 4 \Lambda$ the corresponding equations of motion \eqref{2.9} and \eqref{2.10} give two interesting background solutions presented in subsections \ref{sol-1} and \ref{sol-2}.

\subsection{Equations of Motion}

The equations of motion for nonlocal gravity action \eqref{1.2} are derived in \cite{dimitrijevic9} and have the following form:
\begin{align} \label{2.2}
\hat{G}_{\mu\nu} &= G_{\mu\nu} +\Lambda g_{\mu\nu} -\frac{1}{2} g_{\mu\nu} P(R)\, \mathcal{F}(\Box)\, Q(R) + \big(R_{\mu\nu}  - K_{\mu\nu} \big) W +\frac{1}{2} \Omega_{\mu\nu} = 0,
\end{align}
where $\hat{G}_{\mu\nu}$ is nonlocal version of Einstein's tensor, and
\begin{align}
K_{\mu\nu} &= \nabla_\mu \nabla_\nu - g_{\mu\nu}\Box , \label{2.3} \\
  W &= P'(R)\, \mathcal{F}(\Box)\, Q(R) + Q'(R)\, \mathcal{F}(\Box)\, P(R), \label{2.4} \\
  \Omega_{\mu\nu} &= \sum_{n=1}^{\infty} f_n \sum_{\ell=0}^{n-1} S_{\mu\nu}\big(\Box^\ell P(R),\, \Box^{n-1-\ell}\, Q(R)\big) ,
  \label{2.5} \\ S_{\mu\nu}(A,B) &= g_{\mu\nu} \nabla^\alpha A \nabla_\alpha B + g_{\mu\nu} A \Box B -2 \nabla_\mu A \nabla_\nu B  , \label{2.5a}
\end{align}
where $P'(R)$ and $Q'(R)$ are derivatives of $P(R)$ and $Q(R)$ with respect to R, respectively.

From computation in detail, it follows
\begin{equation}
\triangledown^\mu\hat{G}_{\mu\nu} = 0  \, .       \label{2.6}
\end{equation}

Action \eqref{2.1} is particular case of \eqref{1.2} and the corresponding equations of motion for model \eqref{2.1} easily follow from \eqref{2.2}, i.e. equations of motion are
\begin{align}\label{2.7}
&\hat{G}_{\mu\nu} = G_{\mu\nu}+ \Lambda g_{\mu\nu} -\frac{1}{2}g_{\mu\nu} U \mathcal{F}(\Box)U + 2 (R_{\mu\nu}  - \nabla_{\mu}\nabla_{\nu} +   g_{\mu\nu}\Box)\mathcal{F}(\Box) U \nonumber \\
 & + \frac{1}{2}\sum_{n=1}^{+\infty}f_{n}\sum_{\ell=0}^{n-1}\Big
(g_{\mu\nu}(g^{\alpha\beta}\partial_{\alpha}\Box^{\ell}U
\partial_{\beta} \Box^{n-1-\ell}U + \Box^{\ell} U \Box^{n-\ell}U) \nonumber \\
& - 2 \partial_{\mu} \Box^{\ell}U \partial_{\nu}\Box^{n-1-\ell}U \Big ) =  0,
\end{align}
where $G_{\mu\nu} = R_{\mu\nu} - \frac{1}{2} R g_{\mu\nu}$ is the Einstein tensor, and  $U = R - 4 \Lambda$.

In the sequel of this paper, we are mainly interested in finding and investigating some exact cosmological solutions of \eqref{2.7}.  Since the universe is homogeneous and isotropic at large scales, it has the
Friedmann-Lema\^{\i}tre-Robertson-Walker (FLRW) metric
\begin{align}
ds^2 = - dt^2 + a^2(t)\left(\frac{dr^2}{1-k r^2} + r^2 d\theta^2 + r^2 \sin^2 \theta d\phi^2\right), \quad (c=1) , \, \, k= 0, \pm 1 , \label{2.8}
\end{align}
where $a(t)$ is the cosmic scale factor.
 As a consequence of symmetries of the FLRW metric,
 \eqref{2.7} can be reduced to  two independent differential equations and we take trace and 00-component, respectively:
 \begin{align}
&4\Lambda - R  -2 U \, \mathcal{F}(\Box) \, U +  2(R
 + 3 \Box )\,  \mathcal{F}(\Box) \, U  \nonumber \\
 &+ \sum_{n=1}^{+\infty}f_{n}
\sum_{\ell =0}^{n-1} \Big( \partial_{\alpha}\Box^{\ell}\, U \partial^{\alpha} \Box^{n-1-\ell}\, U + 2\Box^{\ell}\, U \Box^{n-\ell}\, U \Big)
  = 0,  \label{2.9}
\end{align}
\begin{align}
 &G_{00} - \Lambda  + \frac{1}{2} U \, \mathcal{F}(\Box) \, U  +  2 (R_{00} - \partial_{0}\partial_{0} - \Box) \,  \mathcal{F}(\Box) \, U \nonumber \\
 &- \frac{1}{2} \sum_{n=1}^{+\infty} f_{n} \sum_{\ell =0}^{n-1}\Big( \partial_{\alpha}\Box^{\ell}\, U
\partial^{\alpha} \Box^{n-1-\ell}\, U + \Box^{\ell}\, U \Box^{n-\ell}\, U \nonumber \\
 &+ 2 \partial_{0} \Box^{\ell}\, U \partial_{0}\Box^{n-1-\ell}\, U \Big )  = 0 ,  \label{2.10}
\end{align}
where
\begin{equation}
   R_{00} = - 3 \frac{\ddot{a}}{a} \, , \qquad   G_{00} = 3 \frac{\dot{a}^2  + k}{a^2} \,  .     \label{2.11}
\end{equation}

Eq. \eqref{2.7} can be rewritten in the form
\begin{equation}
\hat{G}_{\mu\nu} = G_{\mu\nu}+ \Lambda g_{\mu\nu} - 8 \pi G \hat{T}_{\mu\nu} = 0 \, , \label{2.12}
\end{equation}
where $ \hat{T}_{\mu\nu}$ can be regarded as a nonlocal gravity analog of the energy-momentum tensor in Einstein's gravity.
The corresponding Friedmann equations to \eqref{2.12} are
\begin{equation}
\frac{\ddot{a}}{a} = - \frac{4\pi G}{3} (\bar{\rho} + 3 \bar{p}) + \frac{\Lambda}{3} \,,  \quad \frac{\dot{a}^2  + k}{a^2} = \frac{8\pi G}{3} \bar{\rho}
+ \frac{\Lambda}{3} \,,   \label{2.13}
\end{equation}
where $\bar{\rho}$ and $\bar{p}$ play a role of the energy density and pressure of the dark side of the universe, respectively.
 The related equation of state is
\begin{equation}
\bar{p} (t) = \bar{w}(t) \, \bar{\rho} (t) .  \label{2.14}
\end{equation}

\subsection{Ghost-free condition}

The spectrum can be found and a possibility to avoid ghost degrees of freedom can be studied by considering the second variation of the action. This task was accomplished in different settings. In paper \cite{biswas6} it was done for an action which contains our action (\ref{2.1}) as one of the terms. In paper \cite{dimitrijevic4} analogous analysis was performed for generic functions $P(R)$ and $Q(R)$. The generic idea is that certain combinations containing the operator function ${\mathcal F}(\Box)$ form kinetic operators for scalar and tensor propagating degrees of freedom. Consequently such combinations must be equal to an exponent of an entire function. The latter has no zeros on the whole complex plane and as such does not result in poles in propagators yielding no new degrees of freedom. Detailed expressions and all the restrictions can be found in the above mentioned references.

\section{Cosmological Solutions}

Our intention is to obtain some exact cosmological solutions of the equations of motion \eqref{2.9} and \eqref{2.10} in the form  $a(t) = A t^m e^{\gamma \ t^2}$, where $m$ and $\gamma$ are some constants. At the beginning we take $U = R - R_0$ and $k = 0$ in the equations of motion. Thus we have three parameters $m,\ \gamma, \ R_0$ that  have to be determined. We found that for $R_0 = 4\ \Lambda$ there are two pairs of solutions for $m$ and $\gamma$: 1) $m = \frac{1}{2} , \, \, \gamma = \frac{\Lambda}{4}$ and 2) $m = 0 , \, \, \gamma = \Lambda$. These background solutions are presented below.

Recall that  scalar curvature for the FLRW metric \eqref{2.8} is
\begin{align}
R (t) = 6\Big(\frac{\ddot a}{a} + \big(\frac{\dot a}{a}\big)^2 +\frac{k}{a^2}\Big).   \label{3.1}
\end{align}
 The d'Alembert-Beltrami operator $\Box$  acts as $\Box R = - \frac{\partial^2}{\partial t^2} {R}-3 H \frac{\partial}{\partial t} {R} ,$ where $H=  \frac{\dot{a}}{a}$ is the Hubble parameter.

Note that the Minkowski space (a(t) = const.,\ $R = \Lambda = k = 0$) is  a solution of equations of motion \eqref{2.9} and \eqref{2.10}.

In what follows, we will present  and briefly discuss some exact cosmological solutions mainly with $\Lambda \neq 0 .$

\subsection{Cosmological solution $ a (t) = A \, \sqrt{t} \, e^{\frac{\Lambda}{4} t^2}  \,, \,\, k=0 $}
\label{sol-1}

For this solution we have
\begin{equation}
\dot{a}(t) = a(t) \frac{1}{2} \Big(t^{-1} + \Lambda t  \Big) , \quad  \ddot{a}(t) = a(t)
\frac{1}{4} \Big(\Lambda^2 t^2 + 4 \Lambda - t^{-2} \Big),  \label{3.1a}
\end{equation}
and scalar curvature \eqref{3.1} becomes
\begin{equation} \label{3.2}
R(t) = 3 \Lambda (\Lambda t^2 + 3) .
\end{equation}

The Hubble parameter is
\begin{align}
 H(t) = \frac{1}{2} \big(  t^{-1} +  \Lambda t \big) , \label{3.3}
 \end{align}
 and its first part ($ \frac{1}{2t}$) is the same as  for the radiation dominance in Einstein's gravity, while the second term
 ($\frac{\Lambda t}{2}$) can be related to the dark energy generated by cosmological constant $\Lambda$. It is evident that this dark radiation is dominated at the small cosmic times and can be ignored compared to $\Lambda$ term at  large times. At the present cosmic time $t_0 = 13.801 \cdot 10^9$ yr and $\Lambda = 0.98 \cdot 10^{-35}\ \text{s}^{-2}$, both terms in \eqref{3.3} are of the same order of magnitude and $H (t_0) = 100.2 \ \text{km/s/Mpc}$.  This value for the Hubble parameter is  larger than current Planck mission result $H_0 = (67.40 \pm 0.50)$ km/s/Mpc \cite{planck2018}. Hence this cosmological solution may be of interest for the early universe with radiation dominance and for far-future accelerated expansion.

   There is useful equality
\begin{align}
\Box \big(R - 4 \Lambda\big) =  - 3\Lambda \big(R - 4 \Lambda \big)    \label{3.4}
\end{align}
which leads to
\begin{align}
\mathcal{F}(\Box) \, \big(R - 4 \Lambda\big) =  \mathcal{F}\big(- 3 \Lambda\big) \, \big(R - 4 \Lambda\big) .  \label{3.5}
\end{align}
$R_{00}$ and $G_{00}$ are:
\begin{align}
R_{00} = \frac{3}{4} \big( t^{-2} - 4\Lambda - \Lambda^2 t^2 \big) \,, \quad G_{00} =  \frac{3}{4} \big( t^{-1} + \Lambda t \big)^2 .  \label{3.6}
\end{align}

Using equality $\Box (R-4 \Lambda)=-3 \Lambda (R-4 \Lambda)$, the trace equation \eqref{2.9} becomes
\begin{align}\label{trace:R-4Lambda- anzac}
&4 \Lambda -R- 10 \Lambda(R-4 \Lambda)\mathcal{F}(-3 \Lambda)+ \big(-6 \Lambda (R-4 \Lambda)^{2}-\dot{R}^{2}\big)\mathcal{F}'(-3 \Lambda)=0.
\end{align}

The $00$ component of EOM \eqref{2.10} becomes
\begin{align}\label{00eq:R-4Lambda- anzac}
&G_{00}- \Lambda + (R-4 \Lambda)\mathcal{F}(-3 \Lambda)(2 R_{00}+ \frac{1}{2}R+ 4 \Lambda)- 2 \mathcal{F}(-3 \Lambda)\ddot{R}\nonumber \\
&+ \frac{1}{2}\big(3 \Lambda (R-4 \Lambda)^{2}-\dot{R}^{2}\big)\mathcal{F}'(-3 \Lambda)=0.
\end{align}

Substituting scalar curvature $R$ into the trace equation \eqref{trace:R-4Lambda- anzac} we obtain
\begin{align}
&4 \Lambda - 3 \Lambda ( 3+ \Lambda t^{2})-10 \Lambda^{2}(5 + 3 \Lambda t^{2})\mathcal{F}(-3 \Lambda)\nonumber \\
&+ \big(-6 \Lambda^{3}(9 \Lambda^{2}t^{4}+ 30 \Lambda t^{2}+ 25)- 36 \Lambda^{4}t^{2}\big)\mathcal{F}'(-3 \Lambda)=0.
\end{align}

Similarly, the $00$ component of EOM \eqref{00eq:R-4Lambda- anzac} becomes
\begin{align}
&\frac{3(1+\Lambda t^{2})^{2}}{4 t^{2}}- \Lambda + \Lambda (5+ 3 \Lambda t^{2})\big(\frac{3}{2t^{2}}+ \frac{5}{2} \Lambda \big)\mathcal{F}(-3 \Lambda)\nonumber \\
&-12 \Lambda^{2} \mathcal{F}(-3 \Lambda)+ \frac{1}{2}\big( 3 \Lambda^{3}(9 \Lambda^{2}t^{4}+ 30 \Lambda t^{2}+ 25)- 36 \Lambda^{4}t^{2}\big)\mathcal{F}'(-3 \Lambda)=0.
\end{align}

Finally, the solution of equations of motion \eqref{2.9} and \eqref{2.10} requires constraints
\begin{align}
    \mathcal{F} \big(- 3\Lambda \big) = - \frac{1}{10 \Lambda} \,,  \quad   \mathcal{F}' \big(- 3\Lambda \big) = 0 \, , \quad \Lambda \neq 0 ,  \label{3.7}
\end{align}
which are satisfied by nonlocal operator
\begin{align}
\mathcal{F}(\Box) = \frac{\Box}{30\Lambda^2} \exp{\left(\frac{\Box}{3\Lambda} + 1\right)} . \label{3.7a}
\end{align}

From \eqref{2.13} follows
\begin{equation}
\bar{\rho} (t) = \frac{3 t^{-2} + 3\Lambda^2 t^2 + 2 \Lambda}{32 \pi G} \,,    \quad \bar{p}(t) =  \frac{t^{-2} -3 \Lambda^2 t^2 - 6 \Lambda}{32 \pi G}  .   \label{3.1.5a}
\end{equation}

One can easily conclude that
\begin{equation}
\bar{w} = \frac{t^{-2} - 3\Lambda^2 t^2 - 6\Lambda}{3 t^{-2} + 3\Lambda^2 t^2  + 2\Lambda} \to   \begin{cases}
-1 ,  \, \, t \to \infty \\
\frac{1}{3} , \quad  t \to  0 .   \label{3.1.5b}
\end{cases}
\end{equation}
From \eqref{3.1.5b}, we see that parameter $\bar{w}$ behaves: (i)   like 1/3 at early times as for the case of radiation and (ii) like -1 as in the usual prediction for the late times  acceleration with cosmological constant $\Lambda$.

\subsection{Cosmological solution $ a (t) = A  \, e^{\Lambda t^2} \,, \,\, k=0  $}
\label{sol-2}

For this solution we have
\begin{align} \label{3.2.1}
&\dot{a}(t) = a(t)\ 2 \Lambda t , \quad \ddot{a}(t) = a(t)\ 2 \Lambda \big(2 \Lambda t^2 + 1\big) ,  \\
&R(t) = 12 \Lambda \big(4 \Lambda t^2 +1\big) , \quad H(t) =2 \Lambda t , \label{3.2.1a} \\
&R_{00} = -6\Lambda \big(1 + 2\Lambda t^2 \big) \ \quad G_{00} = 12 \Lambda^2 t^2 .
\end{align}
There are  useful equalities:
\begin{align} \label{3.2.2}
\Box (R - 4\Lambda) = - 12 \Lambda (R - 4 \Lambda) , \quad  \mathcal{F}(\Box) (R -4 \Lambda) = \mathcal{F}(-12 \Lambda) (R -4 \Lambda) .
\end{align}

Using equalities  \eqref{3.2.2},      
the trace equation \eqref{2.9} becomes
\begin{align}  \label{3.2.3}
&4 \Lambda -R- 64 \Lambda(R-4 \Lambda)\mathcal{F}(-12 \Lambda)+ \big(-24 \Lambda (R-4 \Lambda)^{2}-\dot{R}^{2}\big)\mathcal{F}'(-12 \Lambda)=0.
\end{align}

The $00$ component of EOM \eqref{2.10} is as follows:
\begin{align}  \label{3.2.4}
&G_{00}- \Lambda + (R-4 \Lambda)\mathcal{F}(-12 \Lambda)(2 R_{00}+ \frac{1}{2}R+ 22 \Lambda)- 2 \mathcal{F}(-12 \Lambda)\ddot{R}\nonumber \\
&+ \frac{1}{2}\big(12 \Lambda (R-4 \Lambda)^{2}-\dot{R}^{2}\big)\mathcal{F}'(-12 \Lambda)=0.
\end{align}

Substituting scalar curvature $R$ \eqref{3.2.1a} into the trace equation \eqref{3.2.3} we obtain
\begin{align} \label{3.2.5}
&4 \Lambda - 12 \Lambda ( 1+ 4 \Lambda t^{2})-512 \Lambda^{2}(1 + 6 \Lambda t^{2})\mathcal{F}(-12 \Lambda) \nonumber \\
&+ \big(-1536 \Lambda^{3}(36 \Lambda^{2}t^{4}+ 12 \Lambda t^{2}+ 1)- 9216 \Lambda^{4}t^{2}\big)\mathcal{F}'(-12 \Lambda)=0.
\end{align}

Similarly, the $00$ component of EOM \eqref{3.2.4} becomes
\begin{align} \label{3.2.6}
&12 \Lambda^{2}t^{2}- \Lambda +8 \Lambda ( 1+ 6 \Lambda t^{2})\big(2(-12 \Lambda^{2}t^{2}-6\Lambda)+ 6 \Lambda (1 + 4 \Lambda t^{2})+ 22 \Lambda \big)\mathcal{F}(-12 \Lambda)\nonumber \\
&-2\mathcal{F}(-12 \Lambda) 96 \Lambda^{2}+ \frac{1}{2}\big( 768 \Lambda^{3}(36 \Lambda^{2}t^{4}+ 12 \Lambda t^{2}+ 1)- 9216 \Lambda^{4}t^{2}\big)\mathcal{F}'(-12 \Lambda)=0.
\end{align}

To be satisfied, equations of motion \eqref{2.9} and \eqref{2.10} imply  conditions
\begin{align} \label{3.2.7}
    \mathcal{F} \big(- 12\Lambda \big) = - \frac{1}{64 \Lambda} \,,  \quad   \mathcal{F}' \big(- 12\Lambda \big) = 0 \, , \quad \Lambda \neq 0 ,
\end{align}
that can be realized by
\begin{align}
\mathcal{F}(\Box) = \frac{\Box}{768\Lambda^2} \exp{\left(\frac{\Box}{12\Lambda} + 1\right)} . \label{3.2.7a}
\end{align}

According to \eqref{2.13} follows
\begin{equation}
\bar{\rho}(t) = \frac{\Lambda \big(12\Lambda t^2 -1\big)}{8\pi G} \,, \quad \bar{p}(t) = - \frac{3\Lambda \big(4 \Lambda t^2 +1   \big)}{8\pi G} . \label{3.2.8}
\end{equation}

The corresponding $\bar{w}$ parameter is
\begin{equation} \label{3.2.9}
\bar{w} =  \frac{ - 12\Lambda t^2 - 3}{ 12\Lambda t^2  - 1} \to  \begin{cases}
-1 ,  \, \, t \to \infty \\
3 , \quad  t \to  0 .
\end{cases}
\end{equation}

\bigskip

\subsection{Other vacuum solutions: $R (t)$ = const.}

The above two cosmological solutions have scalar curvature $R (t)$ dependent on time $t$.
There are also vacuum solutions with  $R = 4 \Lambda$ that are the same as for Einstein's equations of motion.
Since $\Box (R -4 \Lambda) = 0$, it is evident that such solutions satisfy equations of motion \eqref{2.9} and \eqref{2.10}
without conditions of function $\mathcal{F} (\Box)$.

In addition to the already mentioned Minkowski space, there is another solution with $R(t) = 0$:

 Milne solution: $a(t) = t, \ k= -1, \ \Lambda = 0 , \ (c=1).$

From our analysis follows that there are no other exact power law solutions of the form $a(t) = A t^\alpha$ except  this Milne one.

\section{Concluding Remarks}

In this article, we have presented some exact cosmological solutions of nonlocal gravity model without matter given by
action \eqref{2.1}. Two of these solutions are valid only if  $\Lambda \neq 0 $. The solutions $a(t) = A \sqrt{t} e^{\frac{\Lambda}{4} t^2}$ and $a(t) = A e^{\Lambda t^2}$ are not contained in Einstein's gravity with  cosmological constant $\Lambda$. The solution $a(t) = A \sqrt{t} e^{\frac{\Lambda}{4} t^2}$ mimics interference between expansion with radiation $a(t) = A \sqrt{t}$ and a dark energy $a(t) = A e^{\frac{\Lambda}{4} t^2} .$

The solution $a(t) = A e^{\Lambda t^2}$ is a nonsingular bounce one and an even function of cosmic time.
An exact cosmological solution of the type $a(t) = A e^{\alpha\Lambda t^2}$, where $\alpha$ is a number, appears  also
at least in the following two models of \eqref{1.2}: 1) $P(R) = Q (R) = R$  \cite{koshelev1a} (see also \cite{koshelev1b}), and 2) $P(R) = Q (R) = \sqrt{R - 2\Lambda}$ \cite{PLB}. It would be interesting to investigate other possible models with this kind of solution.

With respect to the cosmological solutions $a(t) = A \sqrt{t} e^{\frac{\Lambda}{4} t^2}$ and $a(t) = A e^{\Lambda t^2}$,
the nonlocal analytic operator $\mathcal{F}(\Box)$ is presented by expressions \eqref{3.7a} and
\eqref{3.2.7a}, respectively. Operator $\mathcal{F}(\Box)$ that takes into account both solutions
should have the form $\mathcal{F}(\Box) = a \frac{u}{\Lambda} \exp(b u^3 + c u^2 + d u) ,$  where $a, b, c, d, $ are some definite constants and $u = \Box/\Lambda $ is dimensionless operator. We do not introduce  an additional parameter like mass $M .$

According to our solutions $a(t) = A \sqrt{t} e^{\frac{\Lambda}{4} t^2}$  and $a(t) = A t^{\frac{2}{3}} e^{\frac{\Lambda}{14} t^2}$ \cite{PLB},
 it follows that effects of the dark radiation ($\sqrt{t}$), the dark matter ($t^{\frac{2}{3}}$) and the dark energy ($e^{\alpha\Lambda t^2}$) at the cosmic scale can be generated by suitable nonlocal gravity models. These findings should play
useful role in further research concerning  the universe evolution.



\section*{Funding}

{This research is  partially  supported by the Ministry of Education, Science and Technological Development of  Republic of Serbia, grant No 174012.}




\end{document}